# Progress with the Prime Focus Spectrograph for the Subaru Telescope: a massively multiplexed optical and near-infrared fiber spectrograph

Hajime Sugai*[a], Naoyuki Tamura[a], Hiroshi Karoji[a], Atsushi Shimono[a], Naruhisa Takato[b], Masahiko Kimura[c], Youichi Ohyama[c], Akitoshi Ueda[d], Hrand Aghazarian[e], Marcio Vital de Arruda[f], Robert H. Barkhouser[g], Charles L. Bennett[g], Steve Bickerton[a], Alexandre Bozier[h], David F. Braun[e], Khanh Bui[i], Christopher M. Capocasale[e], Michael A. Carr[j], Bruno Castilho[f], Yin-Chang Chang[c], Hsin-Yo Chen[c], Richard C.Y. Chou[c], Olivia R. Dawson[e], Richard G. Dekany[k], Eric M. Ek[e], Richard S. Ellis[i], Robin J. English[e], Didier Ferrand[h], Décio Ferreira[f], Charles D. Fisher[e], Mirek Golebiowski[h], James E. Gunn[j], Murdock Hart[g], Timothy M. Heckman[g], Paul T. P. Ho[c], Stephen Hope[g], Larry E. Hovland[e], Shu-Fu Hsu[c], Yen-Shan Hu[c], Pin Jie Huang[c], Marc Jaquet[h], Jennifer E. Karr[c], Jason G. Kempenaar[e], Matthew E. King[e], Olivier Le Fèvre[h], David Le Mignant[h], Hung-Hsu Ling[c], Craig Loomis[j], Robert H. Lupton[j], Fabrice Madec[h], Peter Mao[i], Lucas Souza Marrara[f], Brice Ménard[g], Chaz Morantz[e], Hitoshi Murayama[a], Graham J. Murray[l], Antonio Cesar de Oliveira[f], Claudia Mendes de Oliveira[m], Ligia Souza de Oliveira[f], Joe D. Orndorff[g], Rodrigo de Paiva Vilaça[f], Eamon J. Partos[e], Sandrine Pascal[h], Thomas Pegot-Ogier[h], Daniel J. Reiley[i], Reed Riddle[i], Leandro Santos[f], Jesulino Bispo dos Santos[f], Mark A. Schwochert[e], Michael D. Seiffert[e,i], Stephen A. Smee[g], Roger M. Smith[k], Ronald E. Steinkraus[e], Laerte Sodré Jr[m], David N. Spergel[j], Christian Surace[h], Laurence Tresse[h], Clément Vidal[h], Sebastien Vives[h], Shiang-Yu Wang[c], Chih-Yi Wen[c], Amy C. Wu[e], Rosie Wyse[g], Chi-Hung Yan[c]

[a]Kavli Institute for the Physics and Mathematics of the Universe (WPI), The University of Tokyo, 5-1-5, Kashiwanoha, Kashiwa, 277-8583, Japan;
[b]Subaru Telescope, National Astronomical Observatory of Japan, 650 North A`ohoku Pl., Hilo, Hawaii, 96720, USA;
[c]Institute of Astronomy and Astrophysics, Academia Sinica, P.O. Box 23-141, Taipei, Taiwan;
[d]National Astronomical Observatory of Japan, 2-21-1 Osawa, Mitaka, Tokyo 181-8588, Japan;
[e]Jet Propulsion Laboratory, 4800 Oak Grove Drive, Pasadena, CA 91109, USA;
[f]Laboratório Nacional de Astrofisica, MCTI, Rua Estados Unidos, 154, Bairro das Nações, Itajubá, MG, Brazil;
[g]Department of Physics and Astronomy, Johns Hopkins University, 3400 North Charles Street, Baltimore, MD 21218, USA;
[h]Aix Marseille Université, CNRS, LAM (Laboratoire d'Astrophysique de Marseille) UMR 7326, 13388, Marseille, France;
[i]Astronomy Department, California Institute of Technology, 1200 East California Blvd, Pasadena, CA 91125, USA;
[j]Department of Astrophysical Sciences, Princeton University, Princeton, NJ, 08544, USA;
[k]Caltech Optical Observatories, 1201 East California Blvd., Pasadena, CA 91125 USA;
[l]Centre For Advanced Instrumentation, Durham University, Physics Dept, Rochester Bldg, South Road, Durham DH1 3LE, UK;
[m]Instituto de Astronomia, Geofisica e Ciencias Atmosfericas, Universidade de São Paulo, Rua do Matão, 1226 - Cidade Universitária - 05508-090, Brazil

* hajime.sugai@ipmu.jp; phone 81 4 7136-6551; fax 81 4 7136-6576; http://member.ipmu.jp/hajime.sugai/

## ABSTRACT

The Prime Focus Spectrograph (PFS) is an optical/near-infrared multi-fiber spectrograph with 2394 science fibers, which are distributed in 1.3 degree diameter field of view at Subaru 8.2-meter telescope. The simultaneous wide wavelength coverage from 0.38 μm to 1.26 μm, with the resolving power of 3000, strengthens its ability to target three main survey programs: cosmology, Galactic archaeology, and galaxy/AGN evolution. A medium resolution mode with resolving power of 5000 for 0.71 μm to 0.89 μm also will be available by simply exchanging dispersers. PFS takes the role for the spectroscopic part of the Subaru Measurement of Images and Redshifts (SuMIRe) project, while Hyper Suprime-Cam (HSC) works on the imaging part. HSC's excellent image qualities have proven the high quality of the Wide Field Corrector (WFC), which PFS shares with HSC. The PFS collaboration has succeeded in the project Preliminary Design Review and is now in a phase of subsystem Critical Design Reviews and construction.

To transform the telescope plus WFC focal ratio, a 3-mm thick broad-band coated microlens is glued to each fiber tip. The microlenses are molded glass, providing uniform lens dimensions and a variety of refractive-index selection. After successful production of mechanical and optical samples, mass production is now complete. Following careful investigations including Focal Ratio Degradation (FRD) measurements, a higher transmission fiber is selected for the longest part of cable system, while one with a better FRD performance is selected for the fiber-positioner and fiber-slit components, given the more frequent fiber movements and tightly curved structure. Each Fiber positioner consists of two stages of piezo-electric rotary motors. Its engineering model has been produced and tested. After evaluating the statistics of positioning accuracies, collision avoidance software, and interferences (if any) within/between electronics boards, mass production will commence. Fiber positioning will be performed iteratively by taking an image of artificially back-illuminated fibers with the Metrology camera located in the Cassegrain container. The camera is carefully designed so that fiber position measurements are unaffected by small amounts of high special-frequency inaccuracies in WFC lens surface shapes.

Target light carried through the fiber system reaches one of four identical fast-Schmidt spectrograph modules, each with three arms. All optical glass blanks are now being polished. Prototype VPH gratings have been optically tested. CCD production is complete, with standard fully-depleted CCDs for red arms and more-challenging thinner fully-depleted CCDs with blue-optimized coating for blue arms. The active damping system against cooler vibration has been proven to work as predicted, and spectrographs have been designed to avoid small possible residual resonances.

## 1. INTRODUCTION

The Prime Focus Spectrograph (PFS)[1] planned to be mounted on the Subaru 8.2-meter telescope is an optical/near-infrared multi-fiber spectrograph with 2394 science fibers. The position of each fiber is controlled by a fiber positioner, called "Cobra," consisting of two-staged piezo-electric rotary squiggle motors. Each fiber positioner unit samples a circular patrol region of 9.5 mm diameter at the telescope focal plane, and these positioners are arrayed into a hexagonal pattern with 8-mm central distances between adjacent positioners. The whole array extends to a single hexagonal field of view (FOV) whose effective diameter is 1.3 degree. The hexagonal FOV has an advantage for efficient tiling of survey areas, compared with e.g., a circular FOV. Fibers are divided into four groups, each of which enters one of four spectrograph modules in a slit comprising 600 or 597 science fibers in a single row. Each spectrograph module has three color arms to cover a wide wavelength region from 0.38 μm to 1.26 μm with resolving power $\lambda/\delta\lambda$ of 3000. This low resolution is selected as the PFS's basic mode in order to focus on spectroscopic surveys of fainter galaxies and stars targeting cosmology, Galactic archaeology, and galaxy/AGN evolution. The f-ratio from the telescope Wide Field Corrector (WFC) is transformed by a microlens into a larger value so that difficulties of spectrograph design are eased. Double Schmidt design is used for the spectrograph modules. The modules accept the beam up to an F-ratio of 2.5 even if there is some Focal Ratio Degradation (FRD) through a fiber, and refocuses the fiber image by a camera with the F-ratio of 1.09.

Following the 3rd PFS General Collaboration Meeting held in 2012 August, the PFS specification has been refined in the two respects: the wavelength range is now defined from 0.38 μm to 1.26 μm (instead of 1.3 μm previously), thereby reducing the thermal background without significantly sacrificing the scientific potential. A medium spectral resolution mode with resolving power of 5000 for the red arms (0.71 μm to 0.89 μm only) is also included. This addition particularly benefits important Galactic archaeology studies. This new mode is realized by using a simple grating/grism exchange mechanism without changing any other design elements. The basic characteristics of PFS are summarized in Table 1.

PFS takes the spectroscopic part of the Subaru Measurement of Images and Redshifts (SuMIRe) project, while Hyper Suprime-Cam (HSC) works on the imaging part. The HSC observations started with its first light in 2012 August. Its excellent image quality prove the high quality of the Wide Field Corrector (WFC), which PFS shares with HSC. On 2012 October, the technical team successfully carried out a Requirements Review. In this review, it was confirmed that the team is properly defining system-level requirements and is in process of defining subsystem-level requirements. Furthermore, the PFS collaboration succeeded in the project Preliminary Design Review (PDR) in 2013 February. The review panel, consisting of six experienced external reviewers, recommended that the project proceed to the next phase. The science part of the PDR has been revised and recently published in a survey design paper[2]. PFS is now proceeding to several subsystem Critical Design Reviews prior to instrument construction. Figure 1 shows the PFS integration and test flow. It consists of two components: the prime focus part and the spectrograph part, with the fiber system connecting the two parts. In this conference we have eighteen papers on the PFS instrumentation: instrument overview (the present paper), fiber system[3,4,5,6,7,8], fiber positioner[9] & Prime Focus Instrument[10], spectrograph[11,12,13], dewar & detector[14,15,16,17,18], and metrology camera[19]. In the present paper the instrument overview is provided. Details of individual components are found in the corresponding component-dedicated papers. PFS technical first light is scheduled on Subaru in 2017 with the start of Subaru Strategic Program surveys expected in 2019.

The PFS collaboration, led by Kavli IPMU (WPI) with PI Hitoshi Murayama and PFS project office, consists of Caltech/JPL, Princeton & JHU in the USA, LAM in France, USP/LNA in Brazil, and ASIAA in Taiwan. In 2014 January NAOJ/Subaru formally joined the project, and MPA in Germany has more recently joined.

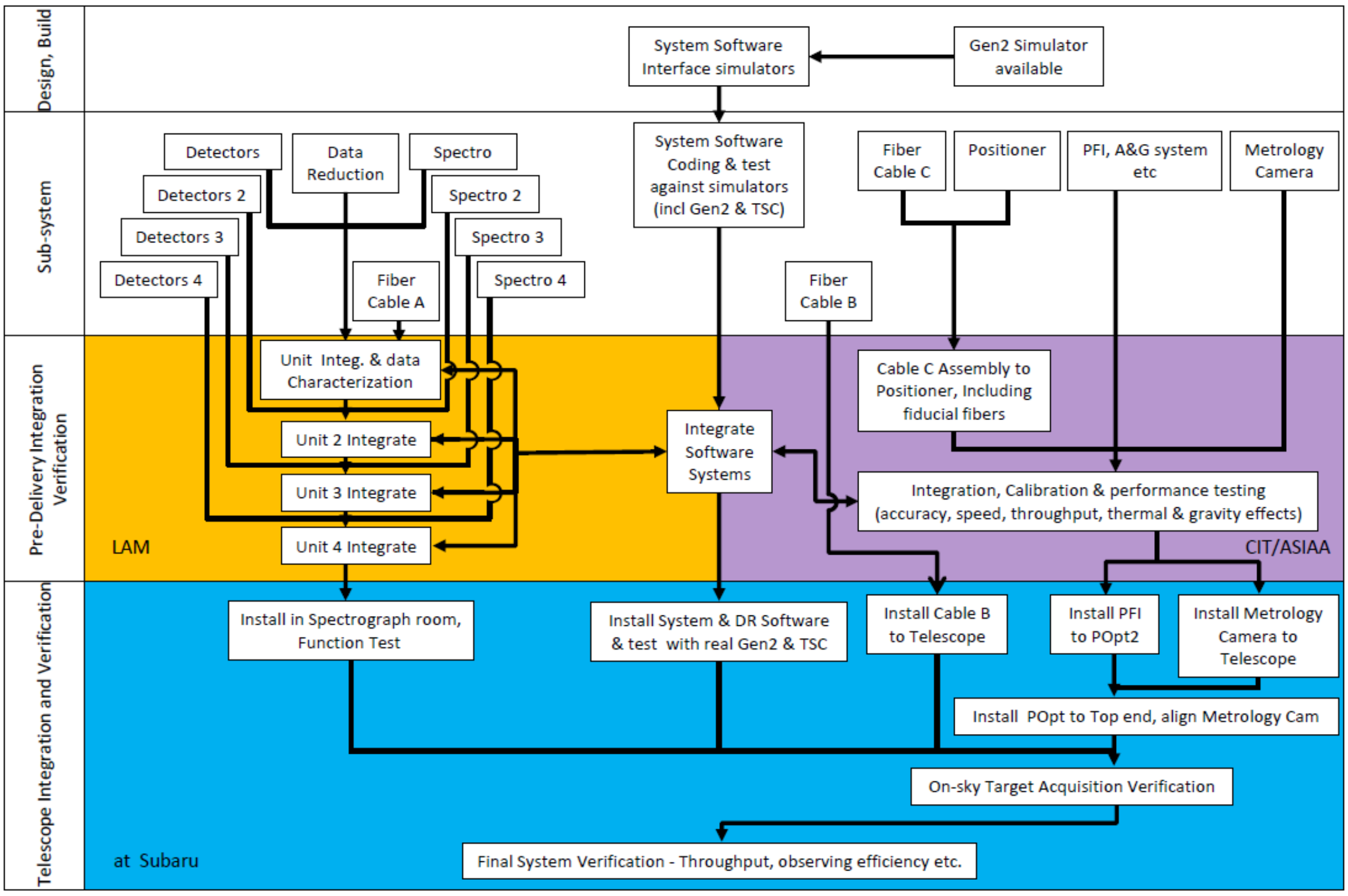

Figure 1. PFS integration and test flow. It consists of two components: Spectrograph and PFI. The fiber system connects these components.

Table 1. Basic characteristics for PFS.

## Basic characteristics for PFS

| | |
|---|---|
| **Field element** | |
| Shape | Plane; thickness 54 mm |
| **Fiber** | |
| Number | 2394 science fibers |
| Diameter (Core/Cladding/Buffer) | Cable C: 127 μm / 165 μm / 189 μm |
| | Core size corresponds to 1”.12 at field center and 1”.02 at corner when a microlens attached |
| | Cable B: 128 μm / 169 μm / 189 μm |
| | Cable A: 129 μm / 168 μm / 189 μm |
| Connectors | Two positions: on telescope spider & on spectrograph benches |
| **Microlens** | |
| Shape | Plano-concave (spherical); thickness 3 mm; glued to fiber input edge |
| F-ratio transformation | F/2.2 to F/2.8 |
| **Fiber positioner** | |
| Positioning mechanism | Two stages of piezo-electric rotary squiggle motors |
| Positioner distribution | Hexagonal pattern |
| Distance from neighboring positioners | 8.0 mm |
| Patrol region | 9.5 mm diameter circle for each fiber |
| **Field shape & size** | |
| Field shape | Hexagon |
| Field size | Diagonal line length |
| | i) Hexagon defined by fiber positioner centers (i.e., twice of distance from field center to farthest fiber positioner centers): 448.00 mm = 1.366 deg on sky |
| | ii) Hexagonal patrol region within which any astronomical target can be accessed at least with one fiber: 453.92 mm = 1.383 deg on sky |
| | Effective diameter of circle whose area is equal to |
| | i) hexagon defined based on fiber positioner centers: 407.41 mm = 1.248 deg on sky |
| | ii) hexagon defined based on patrol region: 412.83 mm = 1.264 deg on sky |
| **Spectrograph** | |
| Number | 4 spectrograph modules, |
| | each with a slit of 600 or 597 science fibers & 3-color arms: located on fourth floor infrared side |
| Slit length | ~140 mm, with center-to-center fiber spacing of 213.93 μm |
| F ratios | All-Schmidt type: collimator F/2.5 & camera F/1.09 |
| Grating | VPH; diameter 280 mm |
| Wavelength region | 380-1260 nm (blue: 380-650 nm; red: 630-970 nm; NIR 940-1260 nm) |
| Spectral resolution | ~2.7 A (~1.6 A for red-arm medium resolution mode in 710-885 nm) |
| **Dewar & Detector** | |
| Dewar window | Camera Schmidt corrector |
| Pixel size | 15 μm |
| Detector | A pair of 2K x 4K fully depleted CCDs for each of blue & red arms; 4K x 4K HgCdTe (1.7 μm cutoff) for NIR arm |
| **Metroloy camera** | |
| Location | At Cassegrain |
| Magnification | 0.0366 |
| Camera aperture size | 380 mm |
| Detector | 50M 3.2μm-pixel CMOS sensor |

## 2. PRESENT STATUS OF DESIGN AND PRODUCTION

### 2.1 Microlens

A glass-molded microlens, produced by Panasonic Industrial Devices Nitto, transforms the f-ratio 2.2 of Subaru telescope plus WFC at prime focus into a slow f-ratio of 2.8. This eases the spectrograph design and also ensures the efficient light acceptance of fiber. Glass molding provides more flexibility of selection of refractive indices compared with plastic molding, and molding technique provides better homogeneity of mechanical dimensions among microlenses compared with polishing. We use high refractive index glass K-VC82 ($n_d$=1.75550) for producing the 3-mm thickness microlenses with 4.8643-mm curvature-radius concave at the entrance surface (Figure 2). A thicker microlens requires a larger curvature radius (slower curve) and results in a smaller light loss for an extended target with respect to the fiber acceptance numerical aperture (NA) while it requires a larger clear aperture. Considering also the manufacturing constraints, a relatively large thickness of 3 mm has been selected. The measured thickness uniformity of mass-produced microlenses was within the specification of +/-10 μm. This uniformity is important in terms of focusing alignment during integration and test. Although focus alignments/checks will be undertaken in a later phase, the uniformity makes focus alignments easier since the microlens foci can be mechanically aligned in advance. The measured outer diameter of mass-produced microlenses was 1.486 mm in average (1.482-1.490 mm in specification) with a dispersion of 1.2 μm (1 sigma). We use a black zirconia fiber arm for integrating a fiber, a microlens, and a fiber positioner motor shaft. The uniformity of outer diameter of microlens is important for the accurate alignment between microlens and fiber axes, since the uniformity makes it possible to tighten the specification on the tolerance of the microlens-holding hole diameter in the fiber arm, which will provide us better mechanical control on the alignment. The concave surface will be broad-band coated (< 1 % between 380 nm and 1.26 μm) and overcoated with $MgF_2$. The opposite flat side will be glued to a fiber.

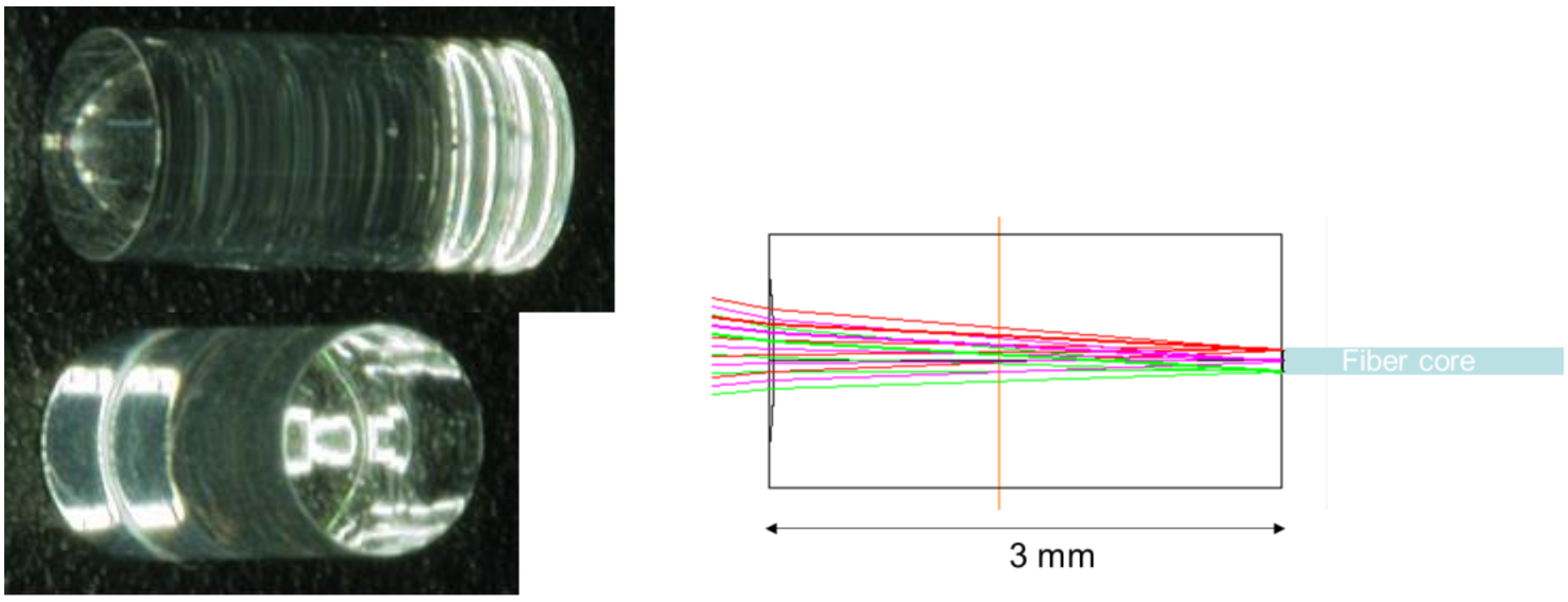

Figure 2. Glass molded microlens will be attached to the entrance of a fiber for F-ratio transformation.

### 2.2 Fiber system

Fiber Connectors provide flexibility in instrument exchange, as well as testing, and retain the possibility of using the fiber system to feed other instruments. The fiber system consists of three components, Cables C, B, and A. These cables are linked by two fiber connectors - one on the telescope spider and one on the spectrograph module benches. On the telescope side, the connectors allow most of the fiber bundle (Cable B) to be permanently installed on the telescope and also facilitate the removal of the Prime Focus Unit, called “POpt2,” as necessary (together with Cable C). At the spectrograph end, the connectors provide easier and safer installation of spectrograph modules (together with Cable A) as well as a way to test the instrument with local sources. They also are important in that a possible independent high resolution spectrograph can be realized relatively easily if desired, making use of the Cobra and all of the fixed fiber system. USCONEC connectors are used, each of which contains 32 holes. Twelve USCONEC connectors are integrated into a single connector, called “Tower” connector between Cables C and B or “Gang” connector between Cables B and A. The Tower connector is thinner/flatter in structure than a cylindrical Gang connector because the former has tighter restrictions in terms of available space volume and shape at the telescope top structure.

Fiber selection has been one of the key issues in the design of the fiber system. Fibers efficiently convey light from the prime focus of the telescope to the spectrograph module inputs along the telescope structure. They need to have high transmission from 380 nm to 1.26μm and to have small Fiber Ratio Degradation (FRD) for this purpose. The project has intensively tested and evaluated particularly fibers that have actually-used diameters and that are produced by two candidate companies: a Japanese company Fujikura and a US company Polymicro. Two methods were used for the measurements of FRD (Figure 3). In the first method, single-wavelength light (filtered for 550 nm or HeNe laser 633 nm) with actually used F-ratio of 2.8 (NA of 0.1786) is input; in the second method, a collimated beam is input and scanned with the incident angle. Both of Fujikura and Polymicro fibers, measured both in 6-meter and 50-meter lengths, showed acceptably small FRDs consistently in the two measuring methods: the estimated losses to the outside of spectrograph acceptance NA corresponding to the F-ratio of 2.5 were 1 to 3 % levels or so. The wavelength dependences of FRD were also investigated and measured to be small enough, δNA of about 0.01 within our whole wavelength range. The absolute transmission measurements for Fujikura 50-meter fibers were carried out and compared with Polymicro case. Results have shown that Fujikura fibers have even better transmission by 2-5 % level for 50-meter length in all the wavelengths we use, and by 10 % level in OH-radical absorption wavelengths such as around 760 nm, 950 nm, and 1.25 μm, although both Fujikura and Polymicro fibers satisfy our specification of spectral transmission.

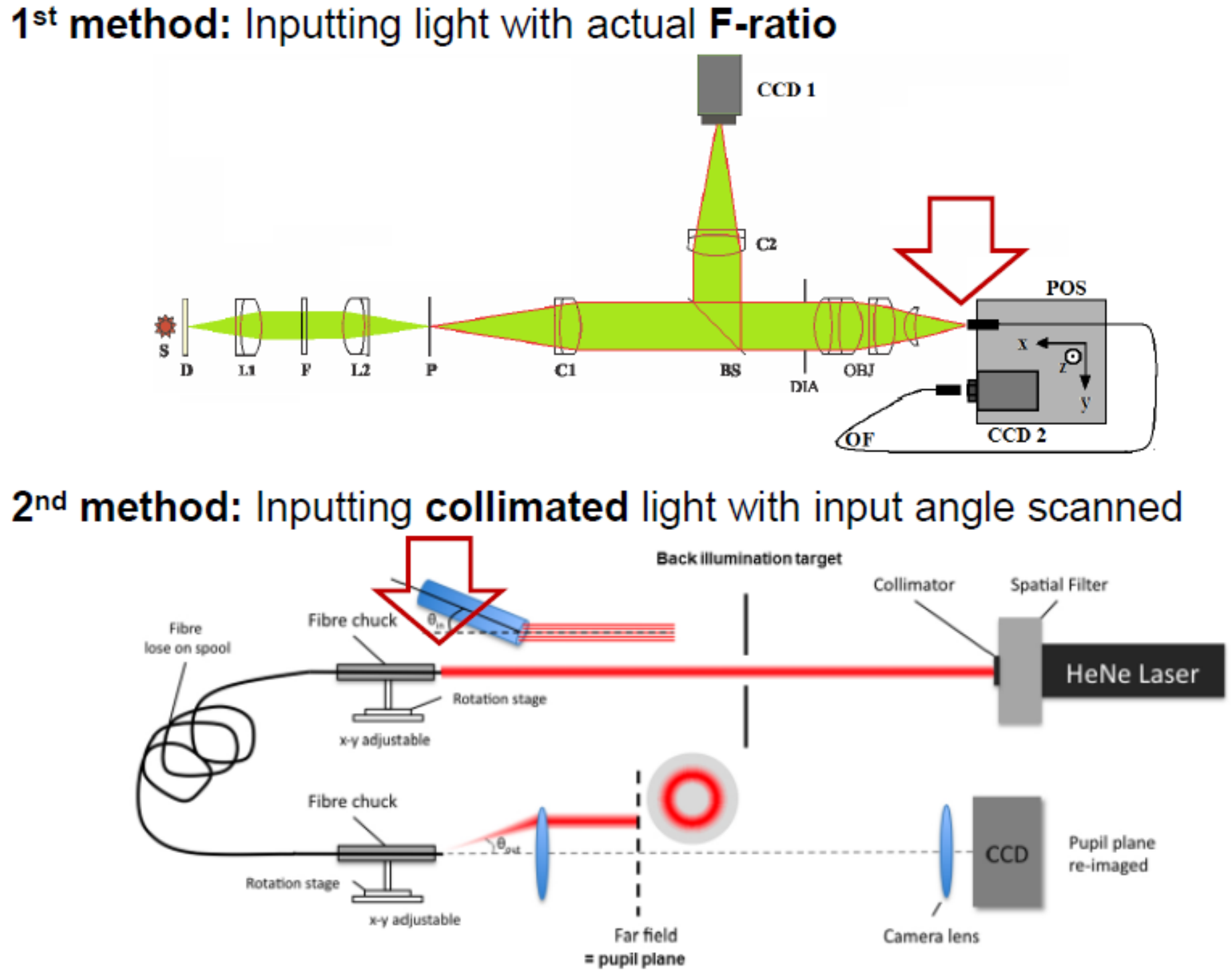

Figure 3. Two methods used for Focal Ratio Degradation (FRD) measurements. In one method, the actually-used F-ratio of 2.8 is input at the fiber entrance and the total effect of FRD is measured. In the other method, collimated light is input and its incident angle is scanned so that the effects on the specific incident angles are measured. When the results from the latter method are integrated in calculation for the beam with the F-ratio 2.8, the result should be consistent with the total FRD measurement in the former method. The upper figure was taken from Cesar et al.'s[20] paper. The measurements made using the latter method were carried out by G. J. Murray, C. N. Dunlop, and J.R. Allington Smith at Durham University CfAI, and the lower figure they kindly provided was originally produced by Ulrike Lemke, who is currently at the Institut für Astrophysik, Göttingen.

Based on the above results, a hybrid type of fiber has been selected: Polymicro fibers for Cables C and A and Fujikura fibers for Cable B. Polymicro has a long history in providing low FRD fibers for astrophysical instrumentation. Our experiments show some possible signature that Polymicro fibers might be slightly better than Fujikura fibers. The project has therefore decided to use Polymicro fibers for Cable C fiber, which has significant moving parts in a small spatial scale when fiber positioners change their configurations, and also for Cable A fiber, which has a small bending radius down to the minimal acceptance of 50 mm at the entrance of the fiber slit to avoid spectrograph beam obscuration. The better transmission of Fujikura fibers has led us to a conclusion to use Fujikura for Cable B, which requires, by far, the longest cable fiber among three cables, about 60 meters.

In order to minimize geometrical losses at the connectors, the outer polyimide-buffer diameter of fibers was specified as 189+/-3 μm for Fujikura and 190+3/-5 μm for Polymicro. This ensures better center alignments between fibers at USCONEC connector, whose hole diameter specification is 195+/-1 μm. To further reduce light loss, we specified core (and clad) diameters as 127 μm+/-3 μm (165+/-3μm), 128 μm+/-5 μm (169+/-2 μm), and 129+/-3 μm (168+/-3 μm), respectively, for Cables C, B, and A. This flow from a smaller core diameter to a larger core diameter (127 < 128 < 129) will minimize geometrical losses in a statistical way, even with slight misalignments of core centers between upstream and downstream fibers, as well as with a possibility of a few μm offset between core center and buffer center. This intentional slight change of core diameters at connectors also will make a slight change in F-ratio., but the effect of geometrical area matching at connectors will be more significant than that of slight F-ratio change. Both of Fujikura and Polymicro fiber productions have already been completed.

To optimize uniformity of fiber image qualities, the mapping of fibers between the fiber positioner and the locations in the fiber slits has been carefully optimized. Field points farther from the center of FOV have worse image properties from the telescope, as well as larger angles of incidence because of non-telecentricity of the WFC. Fibers from these field points should be routed to the inner location of a slit, where the better image quality of the spectrograph is optimized. Conversely, fibers closer to the center of the telescope FOV should be routed to the outer location of the slit. Such a fiber-routing pattern has been successfully designed, maintaining the modularities of fiber positioners and of spectrographs. The slit structure at the entrance of each spectrograph module is based on our new concept of using holes in front- and rear- thin electroformed nickel masks to define the fiber-fiber distances as well as fiber aiming so that the optical axis of each fiber reaches the center of a Kaiser grating, whose clear aperture is technically limited to 280 mm.

### 2.3 Fiber positioner and Prime focus instrument

We use a fiber positioner with two-staged piezo-electric rotary squiggle motors to place a fiber onto a target object. Efforts on reducing its costs as well as on technical improvements were intensively made. Several materials on the end cap of the motor and on the rotor, friction between which creates the rotational motion of the rotor, were investigated; and we have decided to use sapphire rather than expensive machined ceramic on these cost-dominating parts. The possibility of reducing the input voltage to the drive circuits from 100 to 8 volts has also been investigated. Intensive life time tests on such design-modified positioners are being carried out in 300k cycles of clockwise plus counterclockwise rotations including hard stop events and some realistic movements at temperature of -5 degree Celsius, i.e., the most severe environment expected.

Since open-loop positioning accuracy is insufficient, we carry out iterations based on the present position measured in the image of the artificially back-illuminated fiber. Figure 4 shows an example of how the fiber position is converged onto the target position for randomly selected 100 target points within its patrol area of 9.5 mm diameter. Presently 29 Engineering Model (EM) fiber positioners are available and are under the final tests before mass production (Figure 5).

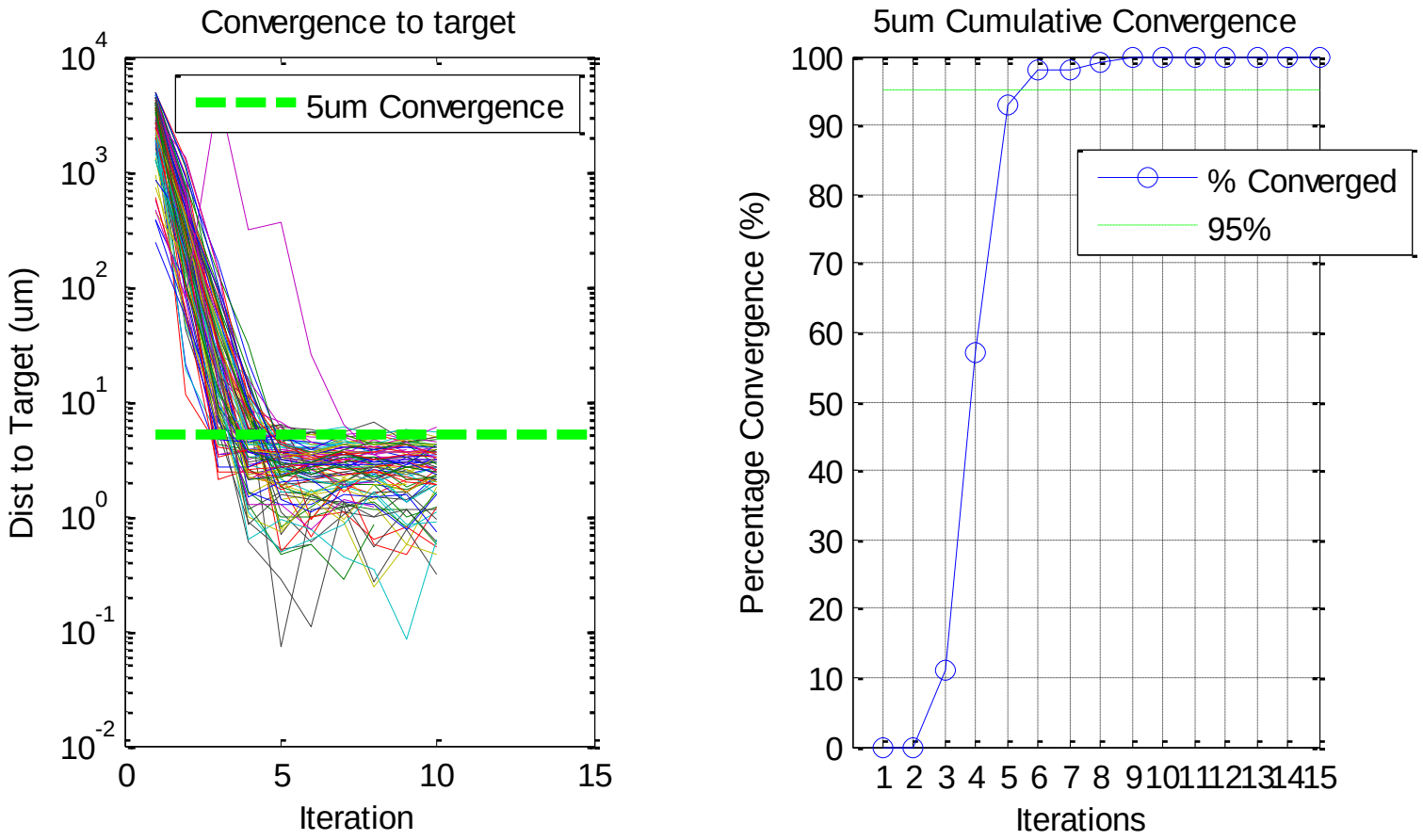

Figure 4. Convergence to target for the best (among five at the time) Engineer Model (EM) fiber positioner. The results shown are for randomly selected 100 target points within the 9.5 mm diameter patrol area. The left figure shows how the actual fiber position is converged with iteration onto the target position for each of these 100 target positions (cases), while the right figure shows the fraction of cases where the fiber positions are converged within 5 μm with respect to the target positions.

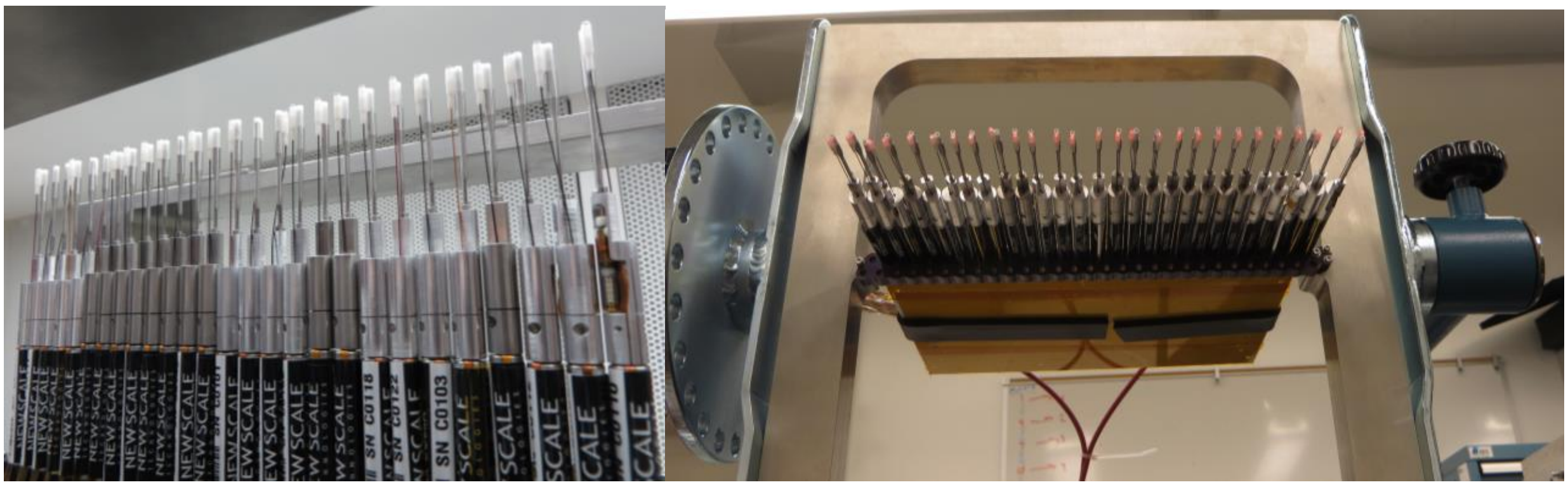

Figure 5. EM fiber positioners. The number of EM positioners is about a half of actually-used number of 57 in two rows of a single module. Many aspects of operating a number of positioners simultaneously will be tested, including such as collisional avoidance between fiber arms and crosstalk check within an electronic board and/or adjacent electronic boards.

42 modules with 57 fiber positioners each are placed on the fiber-positioner optical bench, which is attached to the Prime Focus Instrument (PFI) structure. The PFI is installed in the housing structure called Prime Focus Unit POpt2, whose rotator provides PFI with the rotating part and the non-rotating part with respect to the telescope structure. While the non-rotating part is fixed to the telescope, the rotating enables tracking the target objects in the sky. A structure called positioner frame connects the POpt2 rotator and the fiber-positioner optical bench, on which fiber positioners as well as Acquisition & Guide Camera are mounted. Fiducial fibers are mounted on this optical bench and provide the metrology camera their reference positions when they are artificially back-illuminated. For this purpose, the Fiducial Fiber Illuminator is also located in the PFI rotating part. The Cable Wrapper ensures cables of fibers/electronics/coolant behave in a well-organized way when the rotator moves.

A glass plate called field element is attached to the fiber-positioner optical bench by three field element mounts. The thickness of the field element, 54 mm, roughly corresponds to the total thickness, 52 mm, of a HSC dewar window and a filter that are removed during the instrument exchange from the HSC to the PFS. The slight difference of the thickness is caused by the optimization for the PFS FOV and wavelength range. The field element has an array of thin chrome-deposited dots at the surface closest to microlenses plus fibers. This is for the purpose of better calibration. It is possible to obtain each pure fiber image or each pure spectrum on the spectrograph detectors by hiding some of fibers from calibration sources, such as a flat-field lamp and a wavelength calibration lamp. The dot size has been determined to be about 1.5 mm, considering not only the geometrical obscuration at the distance of 500 μm between the field element and tips of microlenses but also the open-loop accuracy of last fiber positioning. This is because iterations are not possible for hidden fibers and the last movement from the edge of the dot to its center must be sufficiently accurate.

### 2.4 Spectrograph

The optical design of double Schmidt type spectrograph has been completed (Figure 6), and production of optical elements is in progress. All Silica glass blanks are ready and being polished by Winlight System. Figure 7 as an example shows a polished collimator mirror, whose size of 305 mm x 600 mm x 60 mm thickness at its center is the largest in the spectrograph optical elements. The measured curvature radius was 1389 mm (1387.1 +/- 6 mm in specification) and the shape error per each fiber pupil was less than 130 nm peak-to-valley. The roughness was measured as 1 to 1.3 nm root-mean-square. These satisfy our specification for guaranteeing high image quality. Figure 8 shows one of the three prototype Volume Phase Holographic (VPH) gratings produced by Kaiser. Although these are prototypes, they have qualities close to our specification: e.g., the wavefront error measured for the blue VPH grating prototype was 0.4 waves (in 633 nm) root-mean-square for 280 mm clear aperture. Since the trefoil pattern is seen, and if its origin can be identified and corrected, the specified wavefront error will easily be reached for the final gratings. Another possibility of removing this wavefront error is to repolish the surface of the cover plate of the VPH grating in a way that the repolished surface will cancel out the measured trefoil-shaped error pattern. Figure 9 shows an Ohara high-refractive-index prism blank made of S-LAH53 for a medium resolution grism in red channel. Such a blank with high refractive index of $n_d$=1.80610 and with a large size of 320 mm diameter has been successfully produced, realizing the homogeneity of 4 x $10^{-6}$ peak-to-valley for the whole clear aperture when the spherical component is removed. This blank has been cut into two prisms, each of which will be glued to either side of a medium resolution VPH grating to form a medium resolution grism. The exchange between low resolution mode and medium resolution mode is done remotely by using a slide-type exchanger of red gratings/grisms without changing anything else.

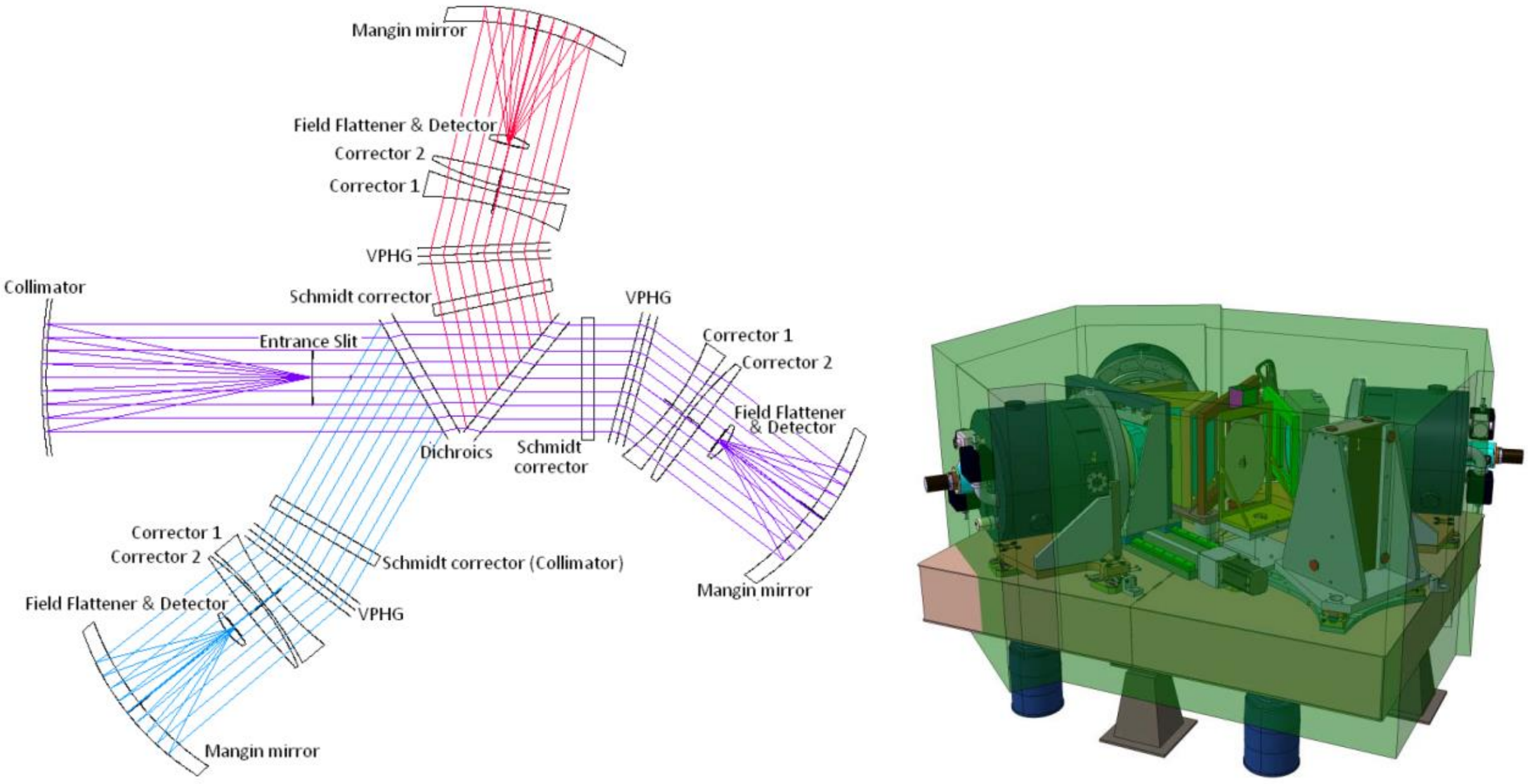

Figure 6. One of four spectograph modules. Schmidt type optics are used both on the f/2.5 collimator and the f/1.09 camera. The entrance slit actually is vertical to this paper and the distance from the slit to the collimator is 693.5 mm. Corrector 1 for each color arm works also as the dewar window. The medium resolution mode for red channel will be realized by using a slide structure (not shown) with a low resolution grating and a medium resolution grism.

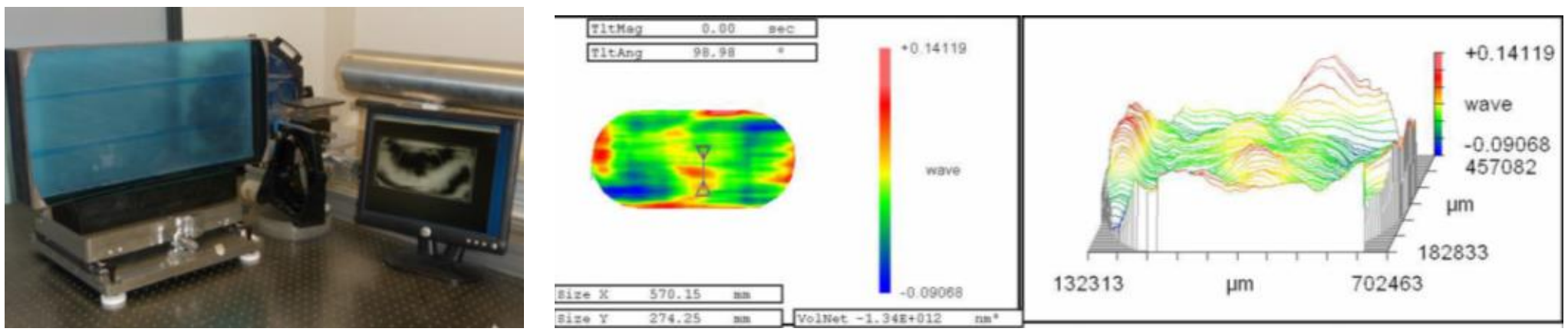

Figure 7. A polished collimator mirror. The size is 305 mm x 600 mm x 60 mm thickness at its center.

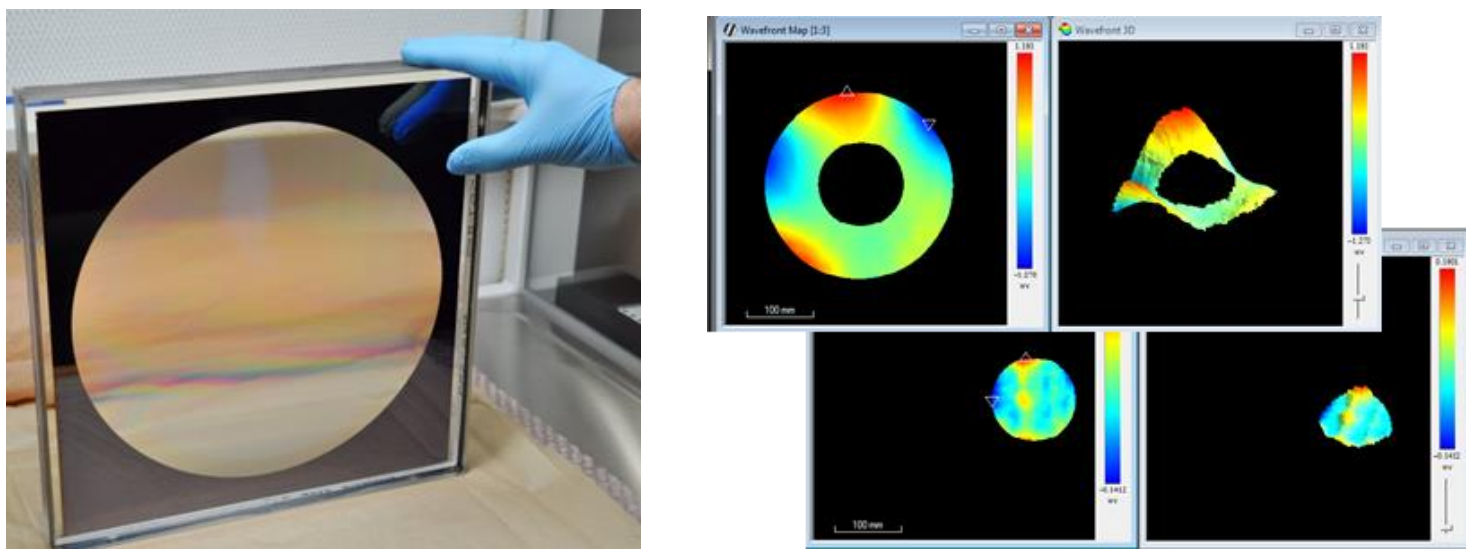

Figure 8. Prototype VPH grating for blue channel. The outer part of grating is masked with Chrome, which provides us the only real light stop in the optical system. The size of the grating is 340 mm x 340 mm x 40 mm thickness.

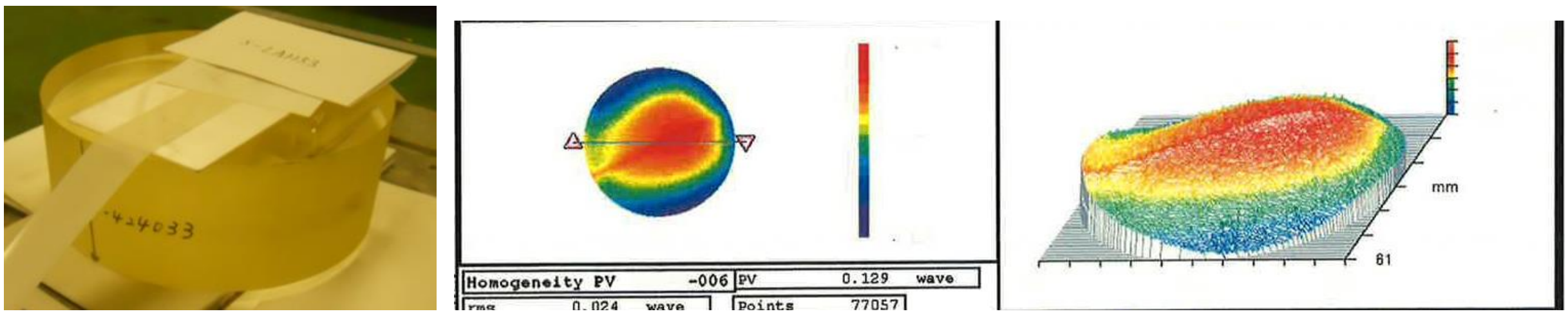

Figure 9. A blank for a pair of prisms that will be attached to a VPH grating to form a red-channel medium -resolution grism. The diameter of the blank is 320 mm.

For the red channel in each spectrograph module, we use a pair of 2K x 4K Hamamatsu fully depleted CCDs with a pixel size of 15 µm x 15 µm. This is exactly the same CCD as used for the HSC. To simplify the control electronics system, mechanical interfaces, as well as operation on optical channels, a pair of 15 µm x 15 µm-pixel 2K x 4K Hamamatsu fully depleted CCDs is also used for blue channel but with newly developed characteristics: thinner fully depletion region and blue optimized coating. The thickness of 100 µm gives us smaller charge diffusion of a sigma of about 4 µm, compared with 7.5 µm for the standard thickness of 200 µm. The blue optimized coating provides us quantum efficiency of 57 %, 74 %, and 91 % at 380 nm, 400 nm, and 500 nm, respectively, which are significantly larger than the case of standard coating. In collaboration with NAOJ Advanced Technology Center (ATC), AlN pin bases were prepared for individual optical CCDs by placing Ti pins with positional accuracy sigma =1.4 µm in each of x, y directions and pin-diameter accuracy of 0.4 µm. These pins allow accurate alignment of a pair of CCDs. For the near-infrared channel in each spectrograph module, a 4K x 4K Teledyne H4RG-15 Mercury-Cadmium Telluride device is used with the same pixel size of 15 µm x 15 µm to the optical CCD case. The special cutoff wavelength of 1.7 µm, as short as feasible, suppresses the thermal background.

The candidate location for setting four spectrograph modules has changed by decision of Subaru observatory from the third floor of the infrared instrument side (IR3) to the fourth floor of the same side (IR4), based on discussion in 2014 January Subaru Users' meeting and on recommendation from Subaru Advisory Committee. The main reasons for this change were: (1) the floor frame structure that has already existed on IR4 requires us much less work/cost of reinforcement, compared with weak structure on IR3; (2) since IR4 floor is farther than IR3 floor with respect to the second floor (Nasmyth), which is crowded with many instruments, the reinforcement of IR4 floor affects to less extent other instruments' operations on the Nasmyth floor; (3) more space is available on IR4 floor compared with IR3 floor with existing structures such as an elevator and beams around the central region of the floor. Although there exists a disadvantage on IR4 in that slightly longer Cable B fibers are necessary (connecting through the second floor, close to the telescope elevation axis) providing some additional light loss, this modification is acceptable (Figure 10).

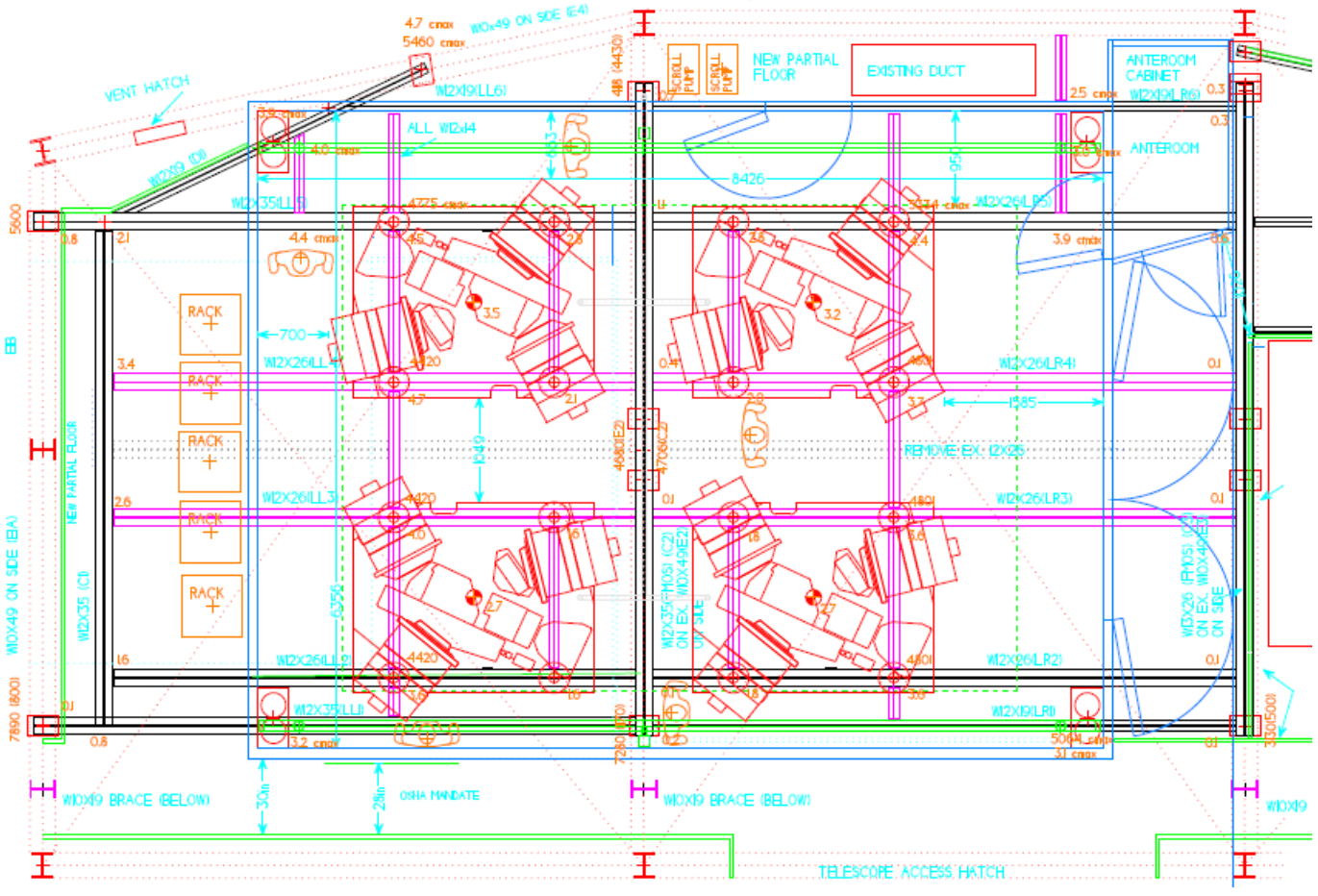

Figure 10. Preliminary floor / spectrograph clean room designs on the fourth floor at infrared side.

Four spectrograph modules are covered with a single light-weighted spectrograph clean room, with thermal insulator walls, which improve both temperature control and cleanliness. The temperature inside of the spectrograph clean room will be kept within +/-1 (+/-0.5 as a goal) degree Celsius variation. This thermal control stabilizes image quality and limits drift of image position to less than 0.2 pixel over 1 hour. The optical bench for each spectrograph module is made of carbon fiber reinforcement plastic (CFRP), which has a low Coefficient of Thermal Expansion, further stabilizing the spectrograph optics configuration against residual temperature variation. The selection of CFRP has also been essential for reducing weight, compared with stainless steel.

The effects of possible vibrations have also been thoroughly investigated. For cooling the insides of dewars as well as detectors in them, we use Sunpower compact free-piston Stirling cryocoolers. These have been selected because of cost and reliability; mean time between failure (MTBF) is about fourteen years. The vibration amplitude produced by the cryocooler is reduced, e.g., down to -40dBG RMS level at the 60 Hz fundamental with an active damper, but the project has carefully designed the mechanical and optical components so that their resonance frequencies are larger than 85 Hz when attached to the optical bench and are larger than 120 Hz for internal modes.

### 2.5 Metrology camera

As described in subsection 2.3, the accuracy of the first positioning of a fiber is insufficient for the desired final positioning accuracy of a small fraction of a fiber core diameter, and therefore it is necessary to measure the present fiber position and iterate the finer positioning based on that information. For this purpose of the positional measurements, we take an image of artificially back-illuminated fibers with a metrology camera located at Cassegrain. The metrology camera is set in the common-use Cassegrain container[21,22], which has been developed in collaboration between Kyoto tridimensional spectrograph II[23,24] team and National Astronomical Observatory of Japan. The container fits with a robotic instrument exchanger Cassegrain Instrument Auto eXchanger for the Subaru Telescope (CIAX)[25,26]. This configuration with a large distance between fibers and the metrology camera allows us to take all fiber images in one shot, although the possible disturbing effects of the air between them merits careful attention. In 2012 December this “dome seeing” effect was measured at Subaru telescope by using the basically same configuration but actually back-illuminated FMOS[27] fibers and a commercial-based CCD camera Atik 450 set at the common-use Cassegrain container (Figure 11). It was concluded that to average out the short-timescale image instability on fiber positions due to the “dome seeing” effects an exposure time of around 1 second or so rather than shorter ones is optimal. The exposure times of e.g., 0.5 second and 1.0 second give us the stability of image position measurements of 2 μm and 2.5 μm in root-mean-square, respectively, at fiber focal plane. These are small enough compared with the final fiber positioning accuracy requirements of an order of 10 μm, a small fraction of PFS fiber core size.

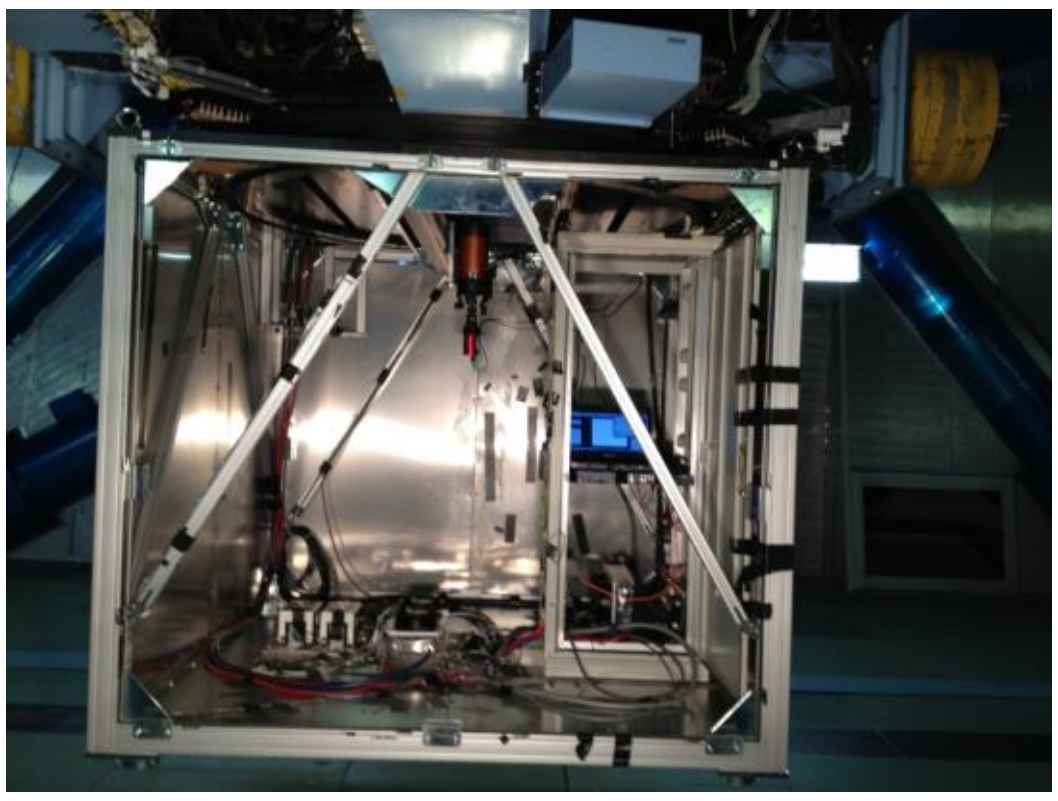

Figure 11. Dome seeing measurements carried out by using FMOS back-illuminated fibers and a commercial-based camera located in the Cassegrain container. The container size is 2000 mm x 2000 mm x 1900 mm height, including the flange to the telescope and short feet used for the setting on CIAX.

The optical path from the fiber plus microlens to the metrology camera is much narrower than the path from the telescope primary mirror to fiber plus microlens, since the metrology camera aperture is much smaller than the primary mirror aperture (Figure 12). This difference does not matter much if the WFC lens surface shapes are perfect, because the different paths share the same chief ray. The WFC lenses, however, have surface shape errors including high spatial frequencies such as 6 mm with the peak-to-valley amplitude of 10-30 nm. Such errors are allowed for the HSC imaging since the amplitude is much less than the wavelengths used. Because the PFS metrology camera however uses only a tiny portion of the WFC lenses for each fiber, the local slopes of the lenses produce significant deformation of coordinates of fibers in the focal plane when they are measured in the metrology camera coordinates. To suppress this effect down to the level of standard deviation of 1.7 μm, an aperture diameter of 380 mm is required for the metrology camera; this diameter corresponds to the unobscured maximum aperture at Cassegrain. The optical and mechanical designs of the metrology camera therefore have been completed for this aperture.

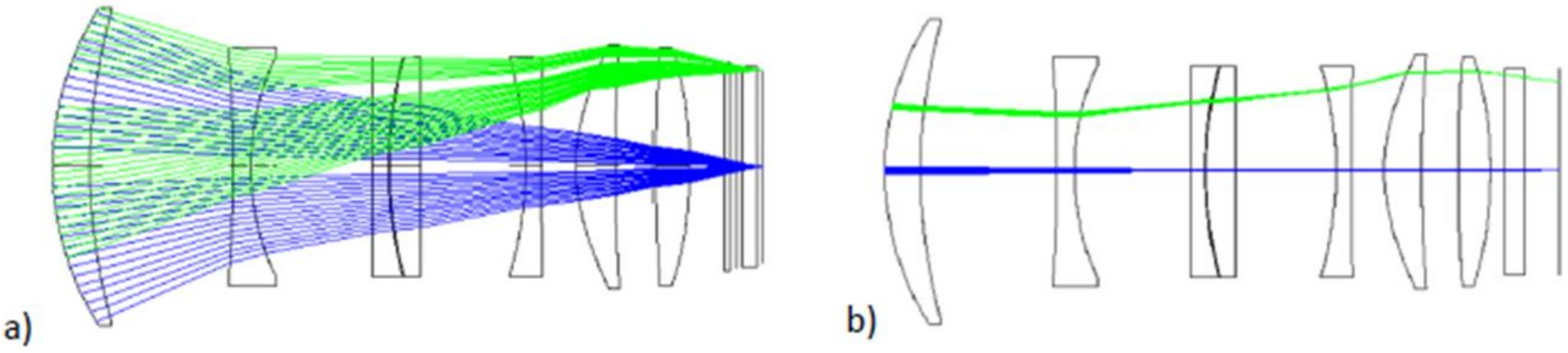

Figure 12. a) Schematic diagram for the light path (coming from the left side of this figure) of a target object through the Subaru telescope. b) Light path from the artificially back-illuminated fibers (at the right side of this figure) to the metrology camera located at Cassegrain. The metrology camera uses a tiny portion of WFC lenses.

## 3. CONCLUSIONS

Upon successful completion of the project PDR, PFS is now in a phase of subsystem Critical Design Reviews and construction, depending on the development phase and required schedule timescale of each subsystem. Mass production of the microlenses is complete, and the fibers have been procured. These two components are now being integrated, which includes gluing them in the fiber arm structure. Later, they will be integrated with a fiber positioner, which is in the final phase before mass production, by gluing the fiber arm to the positioner motor shaft. The positioner modules will be integrated onto the positioner optical bench and then into the PFI structure. This integrated PFI system will be tested together with the metrology camera. For the spectrograph module side, all optical elements, including gratings, are in production. These optical components will be integrated with cryostats and dewars on four spectrograph optical benches. After acceptance tests of each of PFI and spectrograph components, they will be transported to Hawaii, where the whole system finally will be tested all together: PFI system in the POpt2 at the telescope top, spectrograph modules in a single clean room on the fourth floor IR side, and the metrology camera in the Cassegrain container. The technical first light is planned in 2017.

## ACKNOWLEDGEMENTS

We gratefully acknowledge support from the Funding Program for World-Leading Innovative R&D on Science and Technology (FIRST) program “Subaru Measurements of Images and Redshifts (SuMIRe)”, CSTP, Japan, and Fundação de Amparo a Pesquisa do Estado de São Paulo (FAPESP), Brasil. We appreciate staff members at Subaru Telescope for continuously supporting our activities. We thank NAOJ ATC staff members particularly Tetsuo Nishino, Norio Okada, and Yukiko Kamata for preparing AlN pin bases, and Durham University staff members for their consultancy to IPMU on fiber system. We also acknowledge the WFMOS-B team whose accumulated efforts of many years have inspired us.

## REFERENCES

[1] Sugai, H. et al., “Prime Focus Spectrograph – Subaru’s future –,“ Proc. SPIE 8446, 84460Y, 1-13 (2012).
[2] Takada, M. et al., “Extragalactic science, cosmology, and Galactic archaeology with the Subaru Prime Focus Spectrograph,” Publications of the Astronomical Society of Japan 66(1), R1(1-51) (2104).
[3] Takato, N. et al., “Design and performance of a F/#-conversion microlens for Prime Focus Spectrograph at Subaru Telescope,” Proc. SPIE 9147-230, in this conference (2014).
[4] de Oliveira, A. C. et al., “Fiber Optical Cable and Connector System (FOCCoS) for PFS/ Subaru,” Proc. SPIE 9151-168, in this conference (2014).
[5] dos Santos, J. B. et al., “Studying focal ratio degradation of optical fibers with a core size of 128 microns for FOCCoS/ PFS/ Subaru,” Proc. SPIE 9151-189, in this conference (2014).
[6] de Oliveira, A. C. et al., “Slit device for FOCCoS – PFS – Subaru,” Proc. SPIE 9151-165, in this conference (2014).
[7] de Oliveira, A. C. et al., “Polish device for FOCCoS/PFS slit system,” Proc. SPIE 9151-157, in this conference (2014).
[8] de Oliveira, A. C. et al., “Multi-fibers connectors systems for FOCCoS-PFS-Subaru,” Proc. SPIE 9151-229, in this conference (2014).
[9] Fisher, C. D. et al., “Developing Engineering Model Cobra fiber positioners for the Subaru Telescope’s Prime Focus Spectrometer,” Proc. SPIE 9151-68, in this conference (2014).
[10] Wang, S. -Y. et al., “Prime Focus Instrument of Prime Focus Spectrograph for Subaru Telescope,” Proc. SPIE 9147-213, in this conference (2014).
[11] Vivès, S. et al., “Current status of the Spectrograph System for the SuMIRe/PFS,” Proc. SPIE 9147-227, in this conference (2014).
[12] Pascal, S. et al., “Optical design of the SuMiRe/PFS Spectrograph,” Proc. SPIE 9147-158, in this conference (2014).
[13] Madec, F. et al., “Integration and test activities for the SUMIRE Prime Focus Spectrograph at LAM : first results,” Proc. SPIE 9147-222, in this conference (2014).
[14] Hart, M. et al., “Focal Plane Alignment and Detector Characterization for the Subaru Prime Focus Spectrograph,” Proc. SPIE 9154-17, in this conference (2014).
[15] Hope, S. et al., “CCD readout electronics for the Subaru Prime Focus Spectrograph,” Proc. SPIE 9154-88, in this conference (2014).
[16] Hope, S. et al., “Cryocoolervibration damping for the Subaru Prime Focus Spectrograph,” Proc. SPIE 9151-15, in this conference (2014).
[17] Barkhouser, R. H. et al., “VPH gratings for the Subaru PFS: performance measurements of the prototype grating set,” Proc. SPIE 9147-220, in this conference (2014).
[18] Gunn, J. E. et al., “The near infrared camera for the Subaru Prime Focus Spectrograph,” Proc. SPIE 9147-104, in this conference (2014).
[19] Wang, S. -Y. et al., “Metrology camera system of Prime Focus Spectrograph for Subaru telescope,” Proc. SPIE 9147-215, in this conference (2014).

[20] de Oliveira, A. C. et al., "FOCCoS for Subaru PFS" Proc. SPIE 8446, 84464R, 1-14 (2012).
[21] Sugai, H., Hattori, T., Kawai, A., Ozaki, S., Kosugi, G., Ohtani, H., Hayashi, T., Ishigaki, T., Ishii, M., Sasaki, M., Shimono, A., Okita, Y., Sudo, J. and Takeyama, N., "Test Observations of the the Kyoto Tridimensional Spectrograph II at University of Hawaii 88-inch and Subaru Telescopes," Proc. SPIE 5492, 651-660 (2004).
[22] Sugai, H., Ohtani, H., Ozaki, S., Hattori, T., Ishii, M., Ishigaki, T., Hayashi, T., Sasaki, M. and Takeyama, N., "The Kyoto Tridimensional Spectrograph II: Progress," Proc. SPIE 4008, 558-569 (2000).
[23] Sugai, H., Hattori, T., Kawai, A., Ozaki, S., Hayashi, T., Ishigaki, T., Ohtani, H., Shimono, A., Okita, Y., Matubayashi, K., Kosugi, G., Sasaki, M. and Takeyama, N., "The Kyoto Tridimensional Spectrograph II on Subaru and the University of Hawaii 88 in Telescopes," Publications of Astronomical Society of the Pacific 122, 103-118 (2010).
[24] Sugai, H., Ohtani, H., Ishigaki, T., Hayashi, T., Ozaki, S., Hattori, T., Ishii, M., Sasaki, M. and Takeyama, N., "The Kyoto Tridimensional Spectrograph II," Proc. SPIE 3355, 665-674 (1998).
[25] Usuda, T. et al., "CIAX: Cassegrain instrument auto exchanger for the Subaru telescope," Proc. SPIE 4009, 141-150 (2000).
[26] Omata, K. et al., "Control of the Subaru telescope instrument exchanger system," Proc. SPIE 4009, 374-385 (2000).
[27] Kimura M., Maihara T., Iwamuro F., Akiyama M., Tamura N., Dalton G. B., Takato N., Tait P., Ohta K., Eto S., Mochida D., Elms B., Kawate K., Kurakami T., Moritani Y., Noumaru J., Ohshima N., Sumiyoshi M., Yabe K., Brzeski J., Farrell T., Frost G., Gillingham P. R., Haynes R., Moore A.. M, Muller R., Smedley S., Smith G., Bonfield D. G., Brooks C. B., Holmes A. R., Curtis Lake E., Lee H., Lewis I. J., Froud T. R., Tosh I. A., Woodhouse G. F., Blackburn C., Content R., Dipper N., Murray G., Sharples R. and Robertson D. J., "Fibre Multi-Object Spectrograph (FMOS) for the Subaru Telescope," Publications of the Astronomical Society of Japan 62, 1135-1147 (2010).